# Effect of Electrical Properties on Gd modified $BiFeO_3$-$PbZrO_3$

[a]S.K. Satpathy, [a]N.K. Mohanty, [a]A.K Behera, [b]K. C. Patra, [a]Banarji Behera, and [a]P. Nayak

[a]Materials Research Laboratory, School of Physics, Sambalpur University,
Jyoti Vihar, Burla - 768019, Odish, India

[b]Institute of Physics, Bhubaneswar, Odisha

**Abstract**

The $0.5(BiGd_xFe_{1-x}O_3)$-$0.5(PbZrO_3)$ composite was synthesized using a high temperature solid-state reaction technique. Preliminary X-ray structural analysis confirms the formation of the composite. The dielectric constant and loss tangent have been studied. The hysteresis loop suggest that the material is lossy. The impedance parameters were studied using an impedance analyzer in a wide range of frequency ($10^2$-$10^6$ Hz) at different temperatures for all samples. The Nyquist plot suggests the contribution of bulk effect as well as grain boundary effect and the bulk resistance deceases with rise in temperature for all samples. The electrical transport confirms the presence of hopping mechanism in the material. The dc conductivity increases with rise in temperature. The frequency variation of ac conductivity shows that the compound obeys Jonscher's universal power law and confirms the Small Polaron (SP) tunneling effect due to low activation energy for all samples. Temperature dependence of dc and ac conductivity indicates that electrical conduction in the materials are thermally activated process.

**Keywords**: XRD; Dielectric constant; Impedance; relaxation; Electrical conductivity;

**PACS**: 73.61.Ey, 78.20 Ci, 77.80Bh, 72.80.Ng, 84.37. +q

________________

**Corresponding author**: Tel.: +91 663 2431 719

**E-mail**: banarjibehera@gmail.com

1. Introduction

The electrical properties of the composites have recently attracted considerable attention owing to their potential applications i.e., spintronics, high-sensitivity ac magnetic field sensors and electrically tunable microwave devices such as filters, oscillators and phase shifters [1-3]. The composite exhibits both excellent dielectric and soft magnetic properties with a variation of the best candidate for the development of truly integrated passive filters [4]. The activation energy is an important issue on material science. The activation energy is frequency dependent and at higher temperature it may be frequency independent [5]. The sensitivity of a thermistor for temp-sensing is described by the activation energy of the electrical conduction [6]. With the increasing research and application of rare earth elements in composites materials, rare earth doped materials will play an important role in high-tech composite materials. The substitution of rare earth helps to eliminate the impurity phase in materials. $BiFeO_3$ is a rhombohedral crystal structure in the bulk forms [7] and possess a magnetic Neel temperature at 400 °C and a ferroelectric Curie temperature at 820 °C [8, 9]. Further conductivity plays an important role in $BiFeO_3$ [10-13]. Due to high conductivity in $BiFeO_3$ it is possible to develop a MERAM [14] and easily integrated into functional microelectronic devices [15]. Lead zirconate, $PbZrO_3$ (PZO), a perovskite type is the first material which shows antiferroelectricity [16]. Due to its antiferroelectric behavior, PZO is technologically important for applications involving actuators and high energy storage devices [17-22]. The conductivity of $Pb(Zr$ or $Ti)O_3$ has been studied intensively for last few years [23,24]. A new ferroelectric phase was found in $BiFeO_3$-$PbZrO_3$ (BFO-PZO) system, which appears to have the same crystal structure as $PbZrO_3$. Though few works has been done $BiFeO_3$-$PbZrO_3$ [25,26], recently we have reported the electrical properties of $0.5(BiGd_{0.15}Fe_{0.85}O_3)$-$0.5(PbZrO_3)$ and $0.5(BiGd_{0.05}Fe_{0.95}O_3)$-$0.5(PbZrO_3)$ [27,28]. In this work, we elaborately discussed the dielectric, electrical properties and density of state for all concentration of gadolinium doped $BiFeO_3$-$PbZrO_3$.

2. Experimental Details

Gd doped of BFO-PZO with formula $0.5(BiGd_xFe_{1-x}O_3)$-$0.5(PbZrO_3)$ [x = 0.05, 0.10, 0.15, 0.20] were prepared by solid state reaction technique with high purity ingredients, i.e. $Bi_2O_3$, $Gd_2O_3$, $Fe_2O_3$, PbO and $ZrO_2$ in a suitable stoichiometry. The oxides were thoroughly mixed; first in air atmosphere for 2 h, and then in alcohol for 1h. The mixed powders were calcined at an optimized temperature of 750 °C for 5h. The calcined powder was cold pressed into cylindrical pellets of 10 mm diameter and 1-2 mm of thickness at a pressure of $3.5 \times 10^6$ N/m$^2$ using a hydraulic press. PVA (polyvinyl alcohol) was used as binder to reduce the brittleness of the pellet, which was burnt out during the sintering. Then the pellets were sintered at 800°C for 6 h in an air atmosphere. The formation and quality of the compound was studied by an X-ray diffraction (XRD) technique at room temperature with a powder diffractometer (D8 advanced, Bruker, Karmsruhe, Germany) using CuK$_\alpha$ radiation ($\lambda$=1.5405 Å) in a wide range of Bragg's angles 2θ (20°≤2θ≤80°) with a scanning rate of 3°/minute. To study the electrical properties of the composites, the sintered pellets were electroded with air-drying conducting silver paste. After electroding, the pellets were dried at 150 °C for 4 h to remove moisture, if any, and then cooled to room temperature before taking any measurement. The impedance measurements were carried out using a computer-controlled LCR meter (HIOKI Model 3532) in the frequency range of $10^2$-$10^6$ Hz from 25-450 °C.

3. Results and Discussion

Figure 1 shows the X-ray diffraction pattern of 0.5(BiGd$_x$Fe$_{1-x}$O$_3$)-0.5(PbZrO$_3$) [x = 0.05, 0.10, 0.15, 0.20]. The diffraction peaks of the composites were indexed in different crystal systems and unit cell configurations. An orthorhombic unit cell was selected on the basis of good agreement between observed and calculated interplanar spacing d (i.e., $\sum \Delta d = d_{obs} - d_{cal}$ = minimum). The lattice parameters of the selected unit cell were refined using the least-squares sub-routine of a standard computer program package "POWD" [29]. The crystallite size (P) of Gd doped BFO-PZO was roughly estimated from the broadening of a few XRD peaks (in a wide 2θ range) using the Scherrer's equation [30], P=Kλ/($\beta_{1/2}$cosθ$_{hkl}$), (where K=constant=0.89, λ=1.5405 Å and $\beta_{1/2}$ = peak width of the reflection at half intensity). The average values of P were found to be 33 nm for all composition (x = 0.05, 0.10, 0.15 and 0.20). The effect of strain, instruments, and other defects on indexing has been ignored in the calculations.

**3.1 Dielectric Study**

Figure 2. shows, the frequency dependence of dielectric constant at room temperature. At room temperature the dielectric constant increases as the concentration increases but at higher temperature 200 $^0$C (inset) the dielectric constant decreases as the concentration increases and this is a general feature of polar dielectrics. It means that at low dielectric constant show at high frequency region and high dielectric constant at low frequency region, which is exhibiting a typical characteristic of space charge relaxation [31].

Figure 3. shows, the frequency dependence of dielectric loss at room temperature. The loss peak maximum shifts to a lower frequency only due to the effect of polarization by migrating charges at that range. The traditional approach to describing such distorted peaks is to regards them as the superposition of an appropriate number of individual Debye peaks, each occurring at different frequency [32]. It is also observed that at 200 $^0$C (inset) tanδ decreases as frequency increases, which is a typical characteristic of a normal dielectric.

Figure 4. shows, the temperature dependence of dielectric constant and dielectric loss of different compositions at 10 kHz. It is evident that the value of dielectric constant and dielectric loss increases

steadily with increase in temperature for all composition, which is the general feature of a polar dielectrics. At higher temperatures the dielectric constant increases again, which is due to the creation of space charge polarization and conductivity with temperature. It may be due the increase of mobility of space charge carriers with temperature [28].

The hysteresis loop is one of the important criteria for the confirmation of the existence of ferroelectric property in composites. The electric field-induced polarization through (P-E) hysteresis loop for the entire sample at room temperature shown in Figure 5 (a-d). The study revealed that no saturation in polarization-electric field curve could be obtained for the ceramic up to the maximum applied electric field due to the low resistivity for all the composition. The applied electric field beyond 3 kV/cm and further increase in electric field led to electrical breakdown, hence an unsaturated P-E loop for all compositions. For maximum electric field applied to all the sample is 3 kV/cm and hence the remanence polarization ($P_r$) are 0.020, 0.020, 0.015 and 0.015 µC/cm for x=0.05, 0.10, 0.015 and 0.20. The natures of loops suggest that the material is lossy.

## 3.2 Impedance Study

Impedance spectroscopy [33] is determining the relaxation frequency and separation of grain and grain boundary effects, which provides quantitative data of ion transport or ion transport mechanism. The electrical properties are often represented in terms of some complex parameters like complex dielectric constant ($\varepsilon^*$), Complex impedance ($Z^*$), Complex admittance ($Y^*$), Complex modulus ($M^*$) and Loss tangent (tan δ). Electrical ac data of a material are interrelated to each other by the four basic formalisms [34].

$Z^* = Z' - jZ'' = R - j/wC,$     $Y^* = Y' + jY'' = 1/R + jwC,$     $M^* = 1/\varepsilon^* = M' + jM'' = jwC_0Z^*,$     $\varepsilon^* = \varepsilon' - j\varepsilon'',$     $\tan \delta = \varepsilon''/\varepsilon' = M''/M' = -Z'/Z'' = Y'/Y''$

The frequency dependence of imaginary parts represents the capacitive nature. (Shown in figure 6 (a-d).) A peak is observed in the material for all concentration. The peak position regularly shifts towards the higher frequency, which is due to the temperature dependent relaxation process and reduction in the bulk resistivity or may be small polaron hopping with reduction of electron–lattice coupling [35]. The relaxation species may possibly be electrons or immobile species at lower temperature and defects at higher temperature that may be responsible for electrical conduction in the materials [36].

Figure. 7 (a-b) shows the variation between real part of impedance (Z′) with the frequency at 200 $^0$C . Real parts of impedance with imaginary parts of impedance are gives details about the semiconducting behavior. According to Debye's model, a material having a single relaxation time gives rise to ideal semicircles. Figure 7(a). shows one semicircle means only bulk effect and a slight indication of grain boundary effect for $Gd_{0.05}$ and $Gd_{0.10}$. Figure 7(b) a clear indication of both bulk and grain boundary effect in case of $Gd_{0.15}$ and $Gd_{0.20}$.

The equivalent circuit model consists of two parallel RC circuits connected in series. The low frequency region corresponds to grain boundary effect, which is response to $R_{gb}$ and $C_{gb}$ and high frequency region corresponds to bulk effect, which is response to $R_b$ and $C_b$. The equivalent circuit is excellently fitted with R(RC)(RQ) for $Gd_{0.05}$ and $Gd_{0.10}$ and R(C(R(QR)))(CR) for $Gd_{0.15}$ and $Gd_{0.20}$, where R, C and Q are resistance, capacitance and constant phase element (CPE). Admittance of the CPE can be written as: $\gamma_{CPE} = \gamma_0 (j\omega)^n$ where $\gamma_0$ is a constant pre-factor and n the exponent. It is clearly demonstrated that the simple R(CR)(QR) and R(C(R(QR)))(CR) circuit could represent the impedance data ($\chi^2 \sim 10^{-3}$) and the parameters of each fitting are summarized in the table 1.

### 3.3 Conductivity Study:

The electrical conductivity study was performed to incorporate the effect of conduction mechanism and different type of charge carrier in the materials. The value of ac conductivity ($\sigma_{ac}$) of the material was evaluated by using empirical formula $\sigma_{ac} = \varepsilon_0 \varepsilon_r \omega \tan\delta$, where the symbols have their usual meaning.

Figure 8 (a-d). shows the frequency dependence of conductivity for all composition. It is observed that at high frequency region the ac conductivity is frequency independent due to the random diffusion of the ionic charge carriers and the values are corresponds to dc conductivity. The appearance of the dc plateau is an evidence of the formation of conducting path throughout in the material [37]. This can be described by a relation: $\sigma_{ac} = \sigma_o + A\omega^n$ which is Jonscher power law [38], where A is the pre-exponential factor, $\sigma_o$ is the direct current conductivity and n is the fractional exponent ranging 0 to 1 are shown in table 2 . The solid lines in the AC conductivity spectra denote the fit of exponential data to the power law expression: $\sigma_{ac} = \sigma_o + A\omega^n$, which represents the bulk material properties. It is found that Jonsher's power law is best fit to conductivity of the materials at higher frequency for all composition. The deviation from the plateau region of dc conductivity value in lower frequency side of the conductivity spectra is an evidence of the contribution of Electrode Polarization (EP) effect [39]. Generally, for ionic conductors, the power law exponents (n) may lie between 1 and 0.5, representing ideal long-range pathways and diffusion-limited hopping [40]. The term hopping is used to describe the movement of a charge carrier from one site to another. It is observed that the value of n increases linearly with temperature (200-300 $^o$C), which may be a result of the rise of electrode polarization contribution with temperatures for all concentration. In general hopping includes jumps over a potential barrier and quantum mechanical tunneling. Basically n is temperature dependent in Correlated Barrier Hopping (CBH) models, Overlapping Large Polarons (OVL) models and Small Polarons (SP) tunneling models, whereas it is temperature independent in the Quantum

Mechanical Tunneling (QMT) model [41]. figure 9. shown that n is temperature dependent and n is decrease with increases temperature, which follows the small polaron model.

The variation ac conductivity with absolute temperature of the above mention composition is shown in figure 10. at 100 kHz. The conductivity increases as the temperature increases, which is a general behavior of semiconductor. The ac conductivity of disordered solids is at low temperatures and becoming stronger at higher temperatures. The activation energy calculated and is found to be very low (table 3). The low activation energy indicates the hopping conduction mechanism of the materials. The charge carriers involved in the transport are small polarons and these carriers are responsible for the observed conductivity either at high or at low temperatures. Due to the fact that small polaron hopping at low temperatures is due to few-phonon assisted hopping and electrons localized in a disordered materials [42].

The temperature dependence of dc conductivity show in the graph, here all the samples showed a thermally activated process. This can be explained according to the Arrhenius relation:

$$\sigma = \sigma_o \exp(-E_a/k_B T)$$

where $\sigma_o$ the pre-exponential term, $E_a$ is activation energy and $k_B$ is Boltzmann constant. The activation energies are calculated from the slopes of log $\sigma_{dc}$ versus $1000/T$ plots as shown in figure. 11. and it has been found that the activation energy decreases as the concentration decreases. The activation energy (table 4) for different samples was obtained from the least-square straight-line fitting in figure. 11. It is observed that the dc conductivity increases and the activation energy decreases with the increase of Gd concentration. figure. 11. (inset) shows the variation of relaxation time (ln$\tau$) with inverse of absolute temperature ($10^3/T$). The value of $\tau$ decreases with rise in temperature, and thus temperature dependent relaxation time for bulk follows the Arrhenius relation: $\tau = \tau_o \exp(-E_a/k_B T)$ and the activation energy calculated from the least-square straight-line fitting of the data figure. 10 for all composition given in table 4.

4. Coclusion

The composites $0.5(BiGd_xFe_{1-x}O_3)-0.5(PbZrO_3)$ (x = 0.05, 0.10, 0.15, 0.20) were prepared by solid state reaction technique. X-ray analysis exhibits an orthorhombic crystal structure of all the composite at room temperature. The dielectric constant increases as Gd concentration increase. The hysteresis loop suggest the composite is lossy type and the remanent polarization (2P$_r$) found to be 0.040 µC/cm for x = 0.05 and 0.10 and 0.030 µC/cm for x = 0.15 and 0.20. Complex impedance plot reveal the presence of grain and grain boundary effect in the cmposites. The universal jonscher's power law is well fit to the conductivity spectrum of all the composition. The electrical conduction mechanism of the composite may be explained through the small polaron model. The variation of ac and dc conductivity of the material as a function of temperature exhibits Arrhenius type of electrical conductivity.

## Acknowledgement

The authors acknowledge the financial support through DRS-I of UGC under SAP, School of Physics, Sambalpur University. Two of the authors SKS and NKM acknowledge financial support of UGC through UGC-BSR fellowship scheme. One of the author BB acknowledges to the SERB under DST Fast Track Scheme for Young Scientist (Project No. SR/FTP/PS-036/2011) New Delhi, India.) and PN acknowledges CSIR for sanction of Emeritus Scientist scheme (Project No. 21(0944)/12/EMR-II).

\

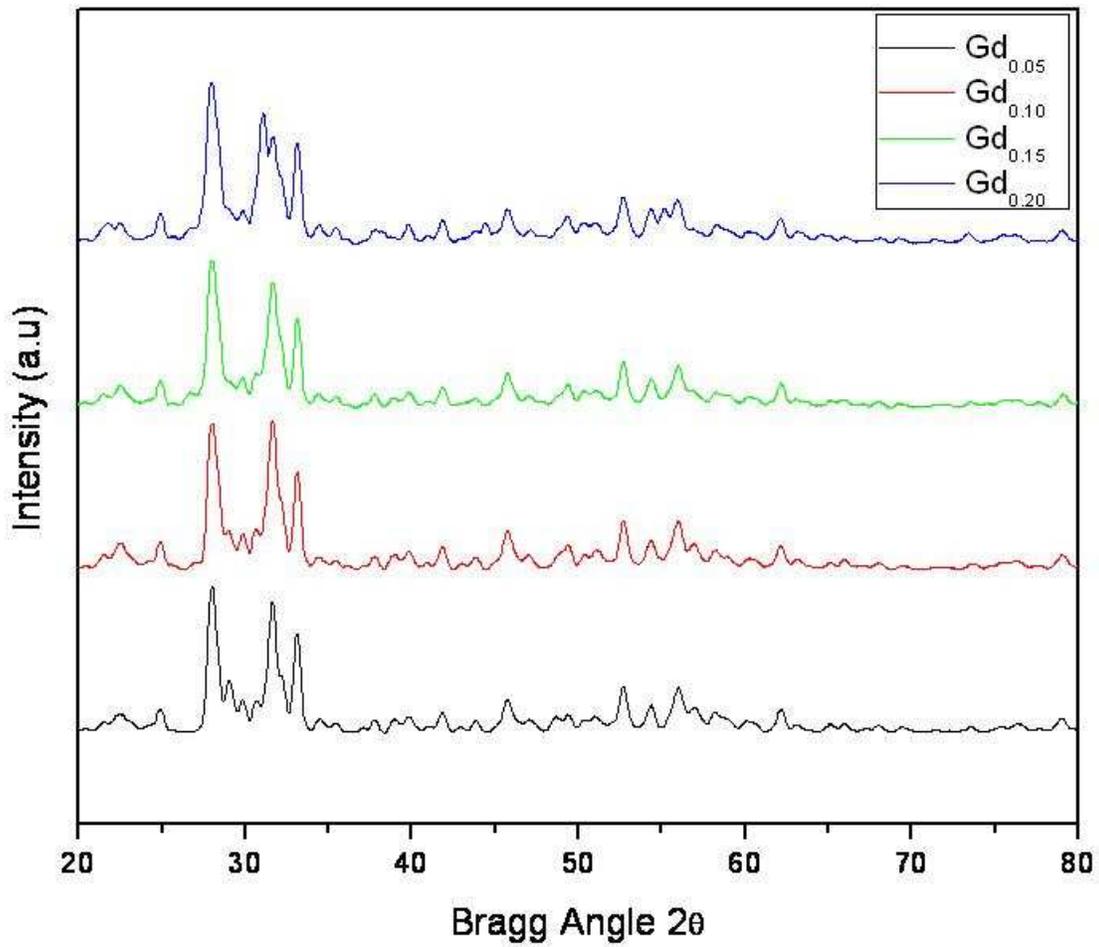

Figure. 1. XRD patterns of $0.5(BiGd_xFe_{1-x}O_3)$-$0.5(PbZrO_3)$ (x=0.05, 0.10, 0.15, 0.20) at room temperature.

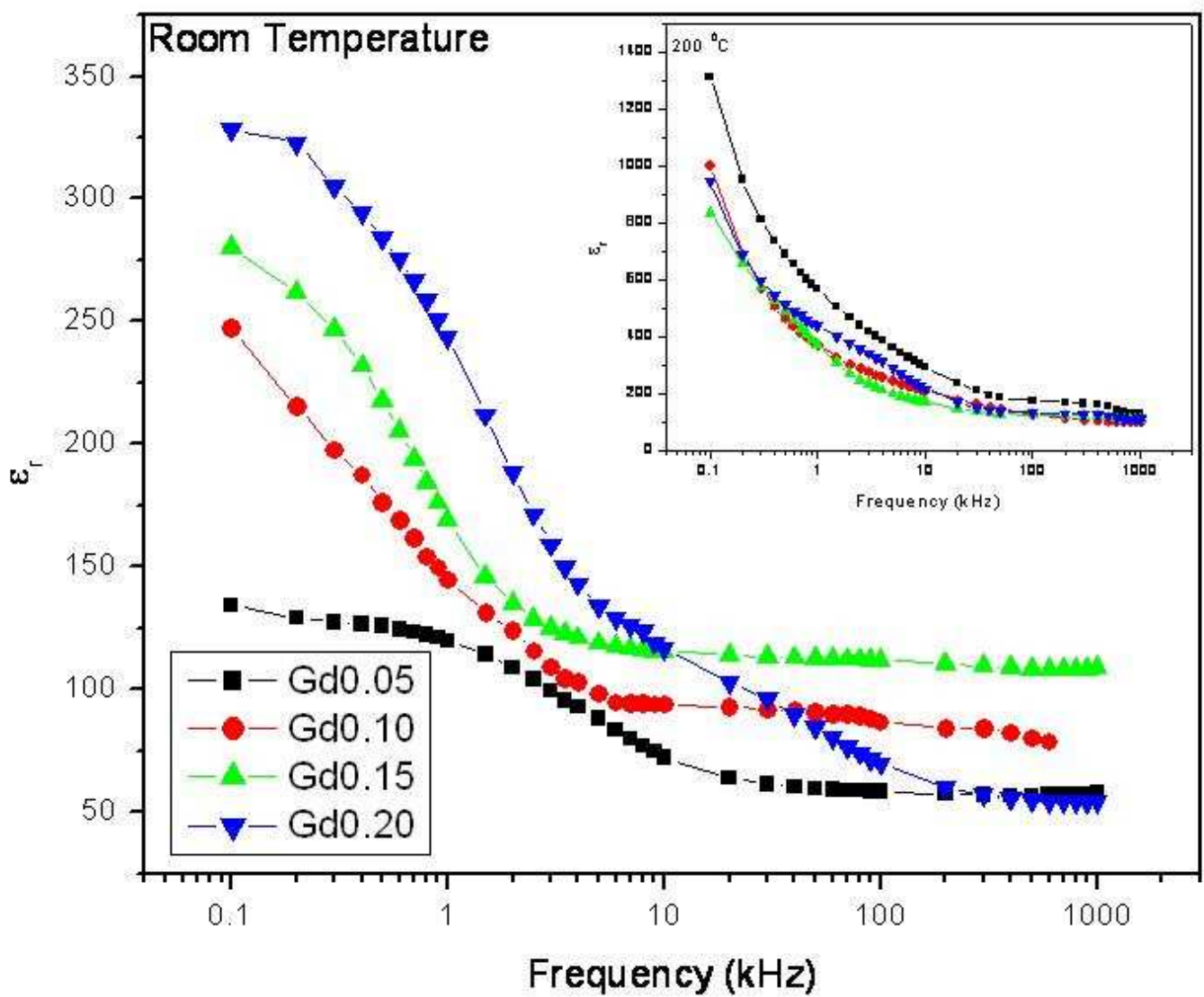

Figure. 2. Variation of relative dielectric constant ($\varepsilon_r$) of $0.5(BiGd_xFe_{1-x}O_3)$-$0.5(PbZrO_3)$ (x=0.05, 0.10, 0.15, 0.20) with frequency at room temperature and 200 $^0$C (inset).

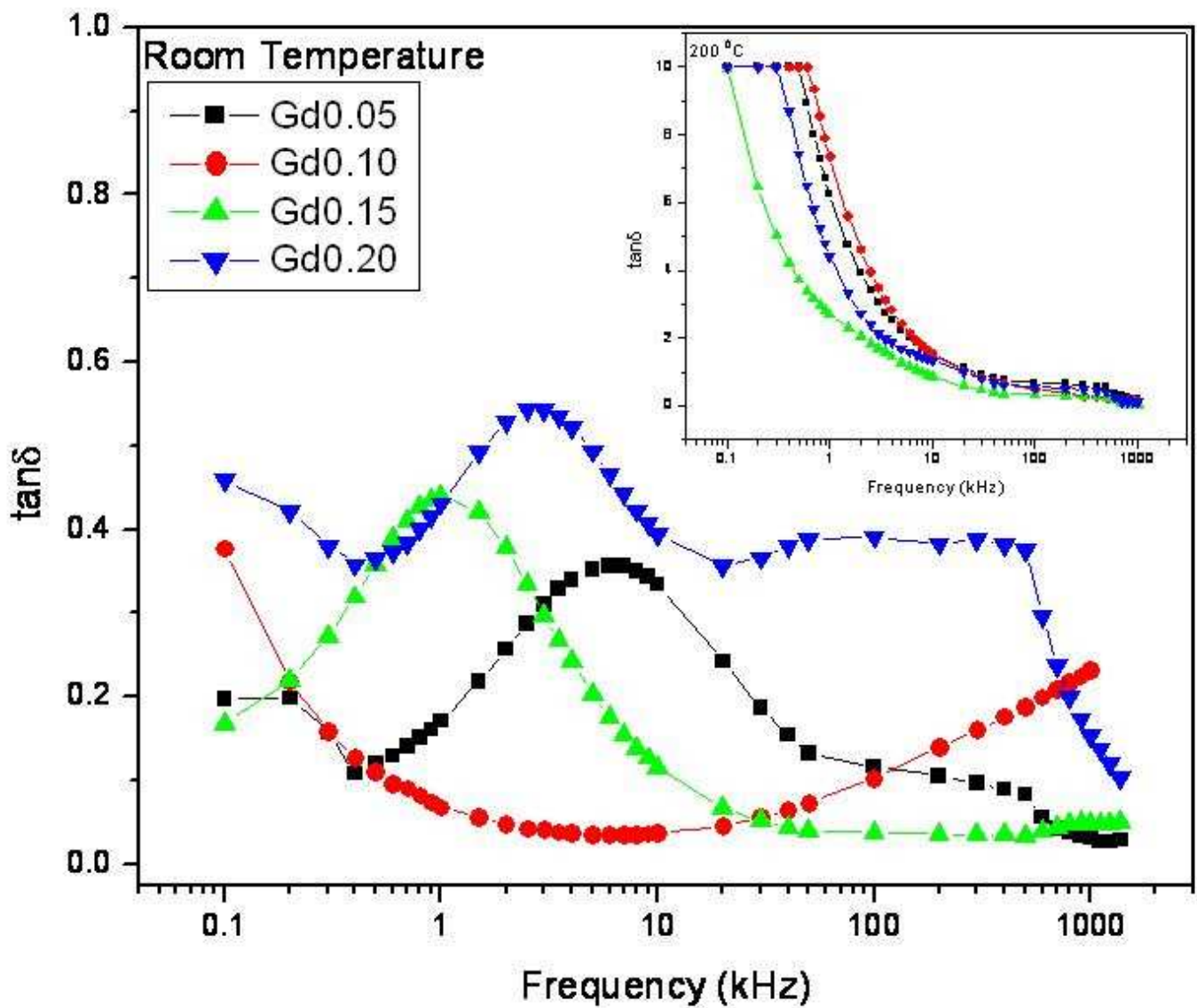

Figure. 3. Variation of loss tangent (tanδ) of $0.5(BiGd_xFe_{1-x}O_3)$-$0.5(PbZrO_3)$ (x=0.05, 0.10, 0.15, 0.20) with frequency at room temperature and 200 $^0$C (inset).

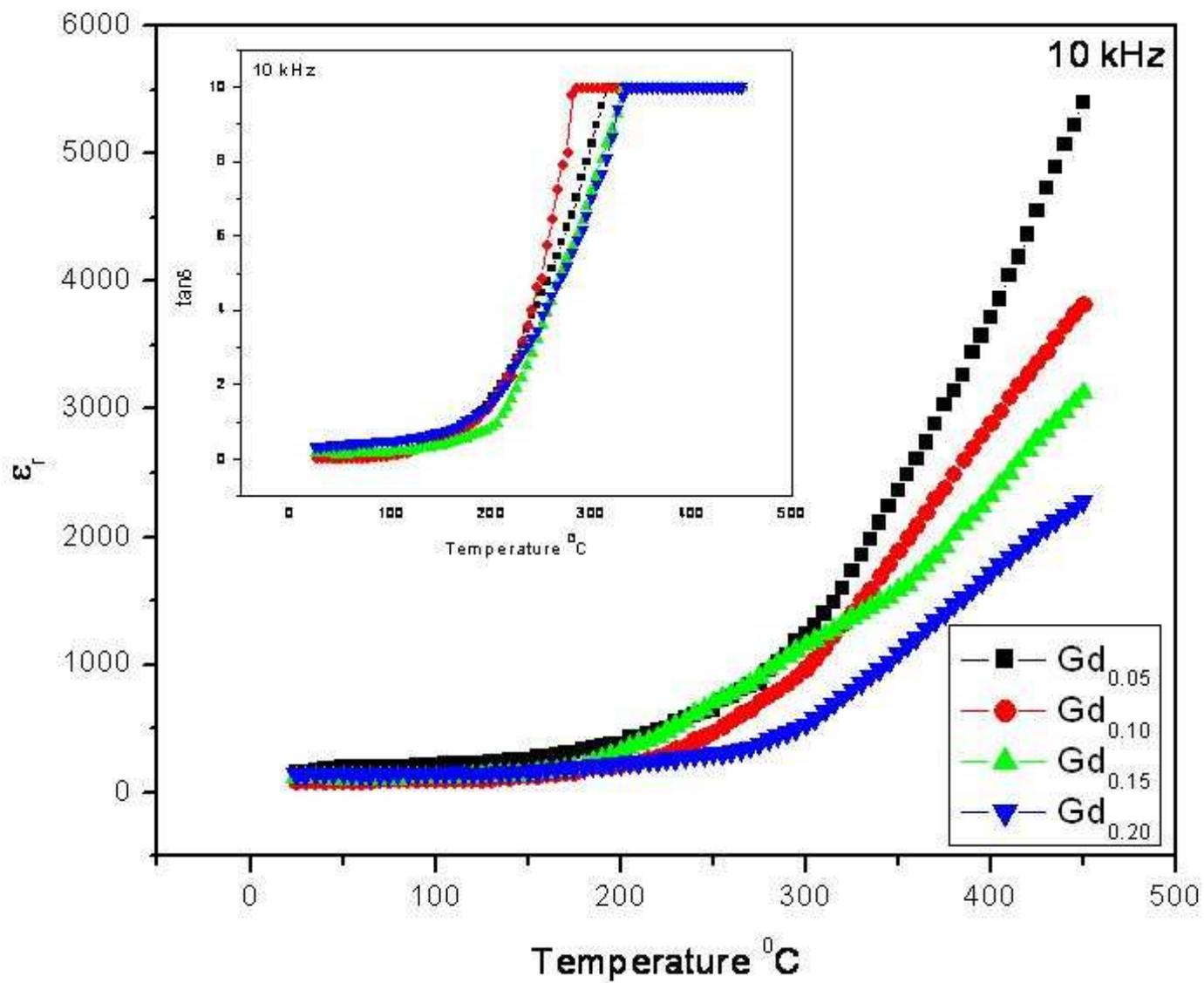

Figure. 4. Variation of relative dielectric constant ($\varepsilon_r$) and loss tangent (tan$\delta$) of 0.5(BiGd$_x$Fe$_{1-x}$O$_3$)-0.5(PbZrO$_3$) (x=0.05, 0.10, 0.15, 0.20) with temperature at 10 kHz.

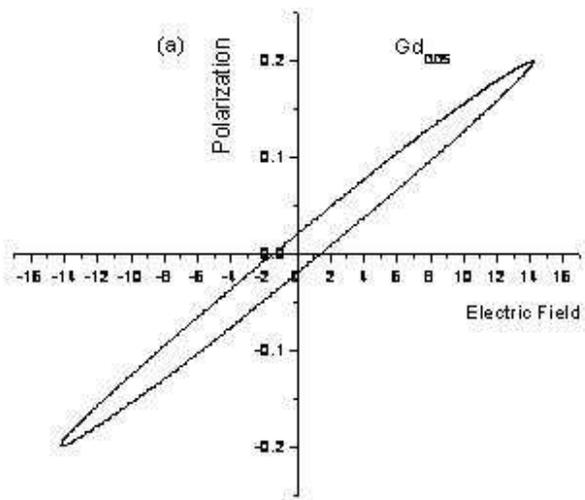
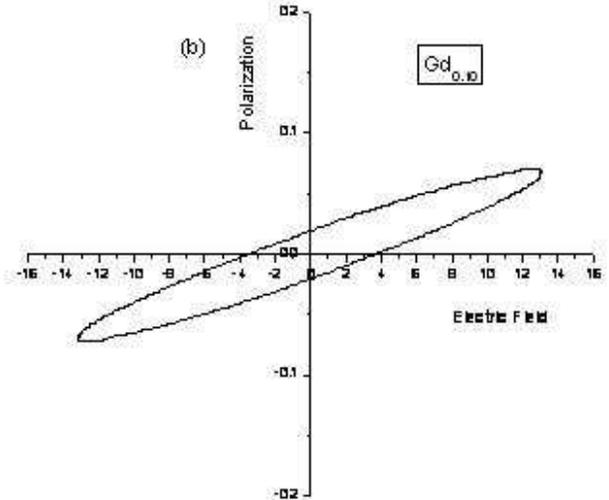
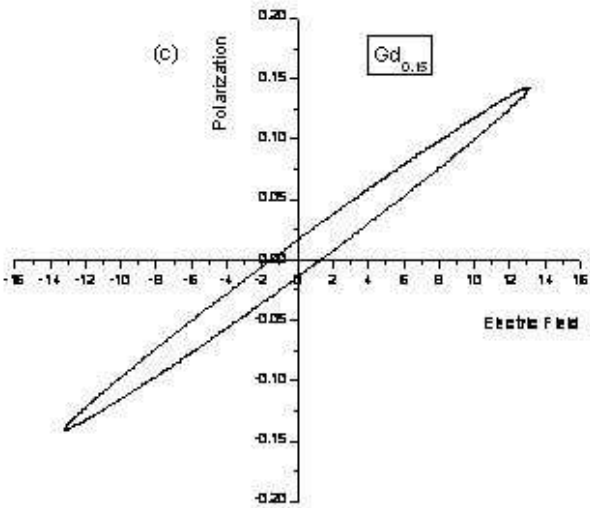
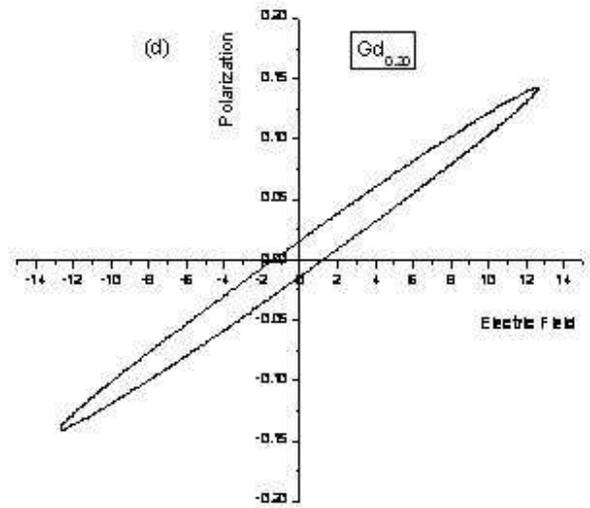

Figure. 5 (a-d) variation of electric field and polarization (P-E loop) of $0.5(BiGd_xFe_{1-x}O_3)$-$0.5(PbZrO_3)$ (x=0.05, 0.10, 0.15, 0.20) at room temperature.

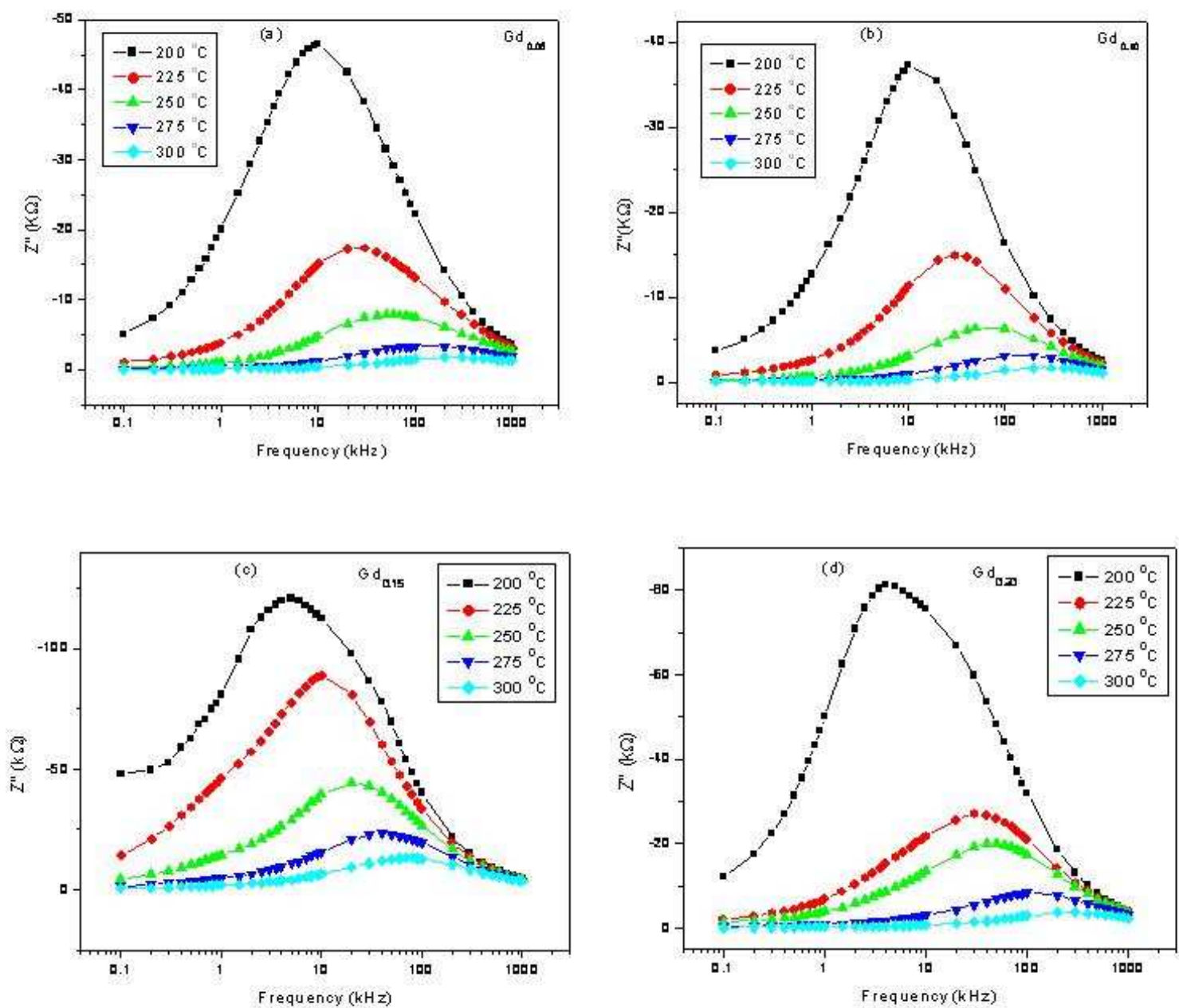

Figure. 6 (a-d). Variation of Z″ as a function of frequency at different temperatures of 0.5(BiGd$_x$Fe$_{1-x}$O$_3$)-0.5(PbZrO$_3$) (x=0.05, 0.10, 0.15, 0.20).

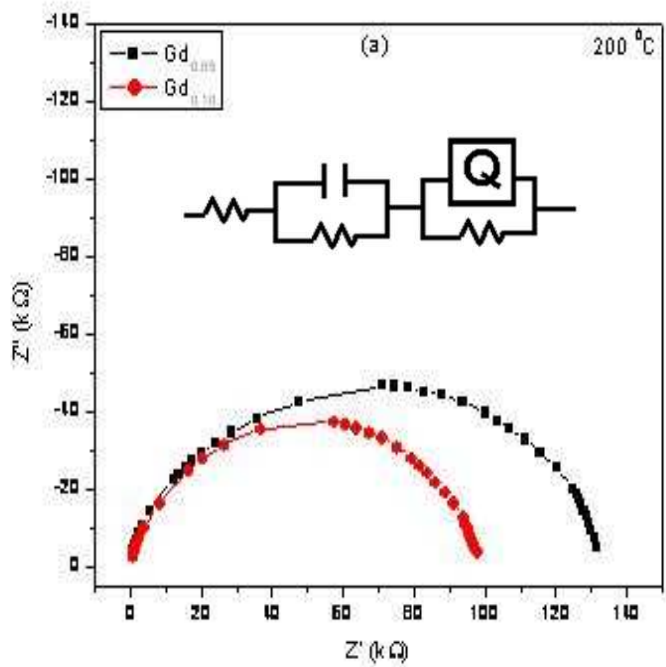 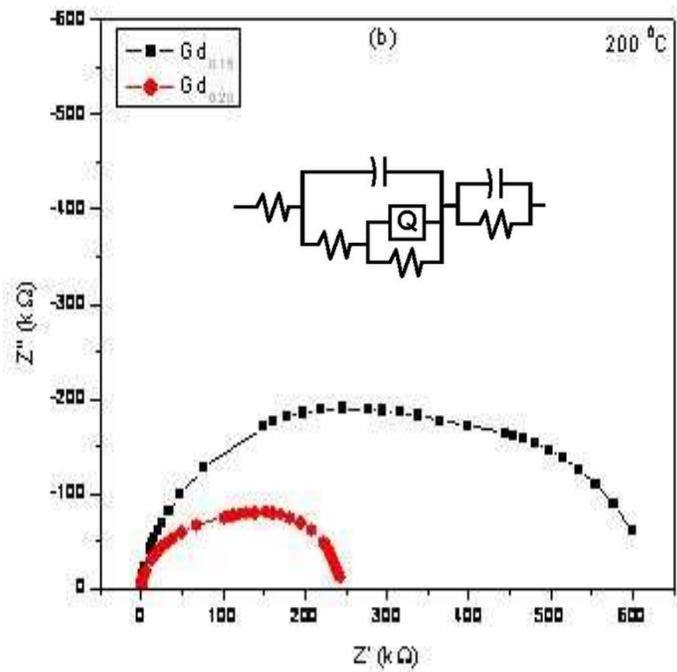

Figure. 7 (a-b). Complex impedance spectrum (Z′ vs. Z″) of $0.5(BiGd_xFe_{1-x}O_3)$-$0.5(PbZrO_3)$ (x=0.05, 0.10, 0.15, 0.20) at 200 $^0$C.

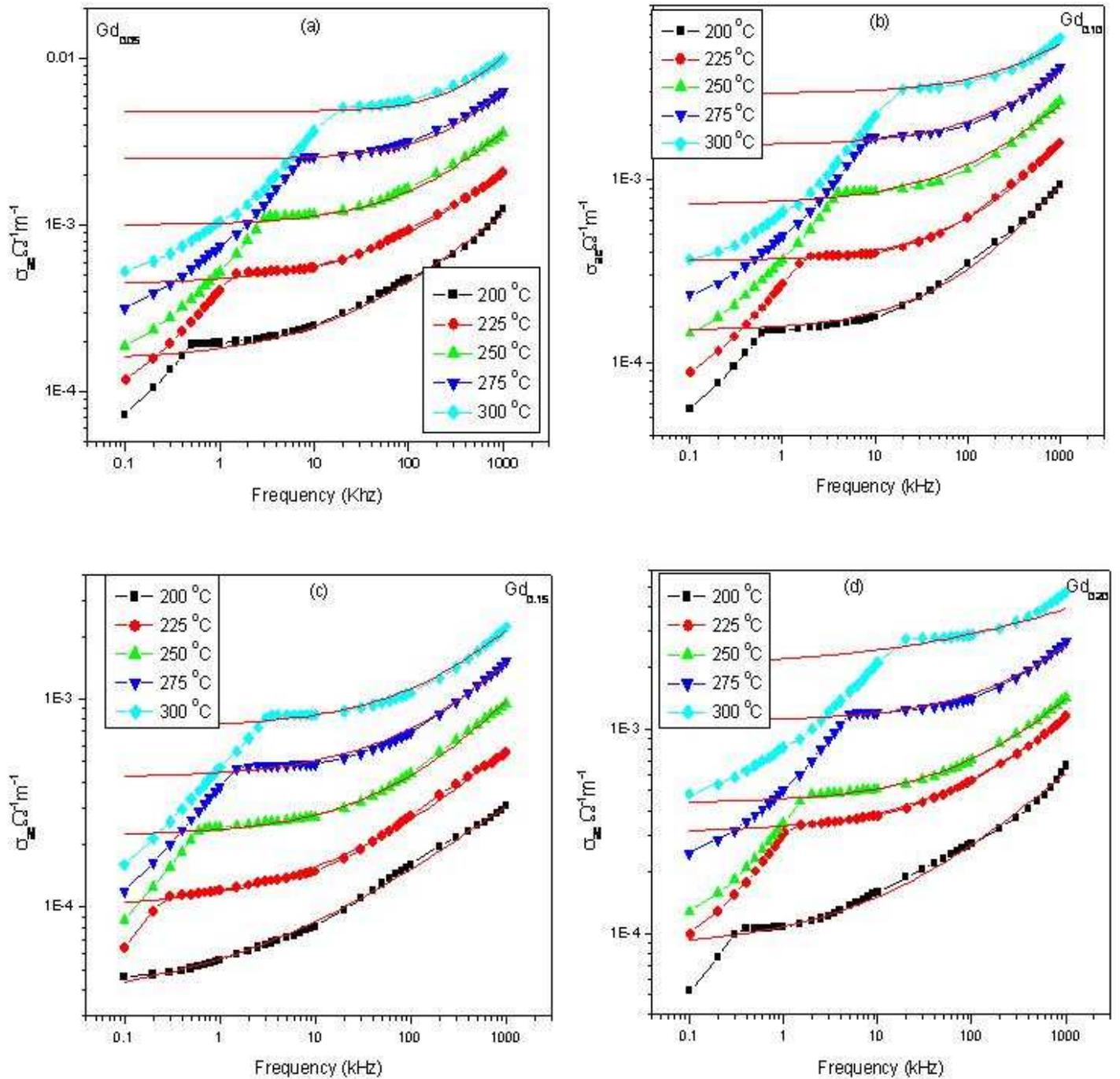

Figure. 8 (a-d). Variation of ac conductivity ($\sigma_{ac}$) of $0.5(BiGd_xFe_{1-x}O_3)$-$0.5(PbZrO_3)$ (x=0.05, 0.10, 0.15, 0.20) with frequency at different temperatures.

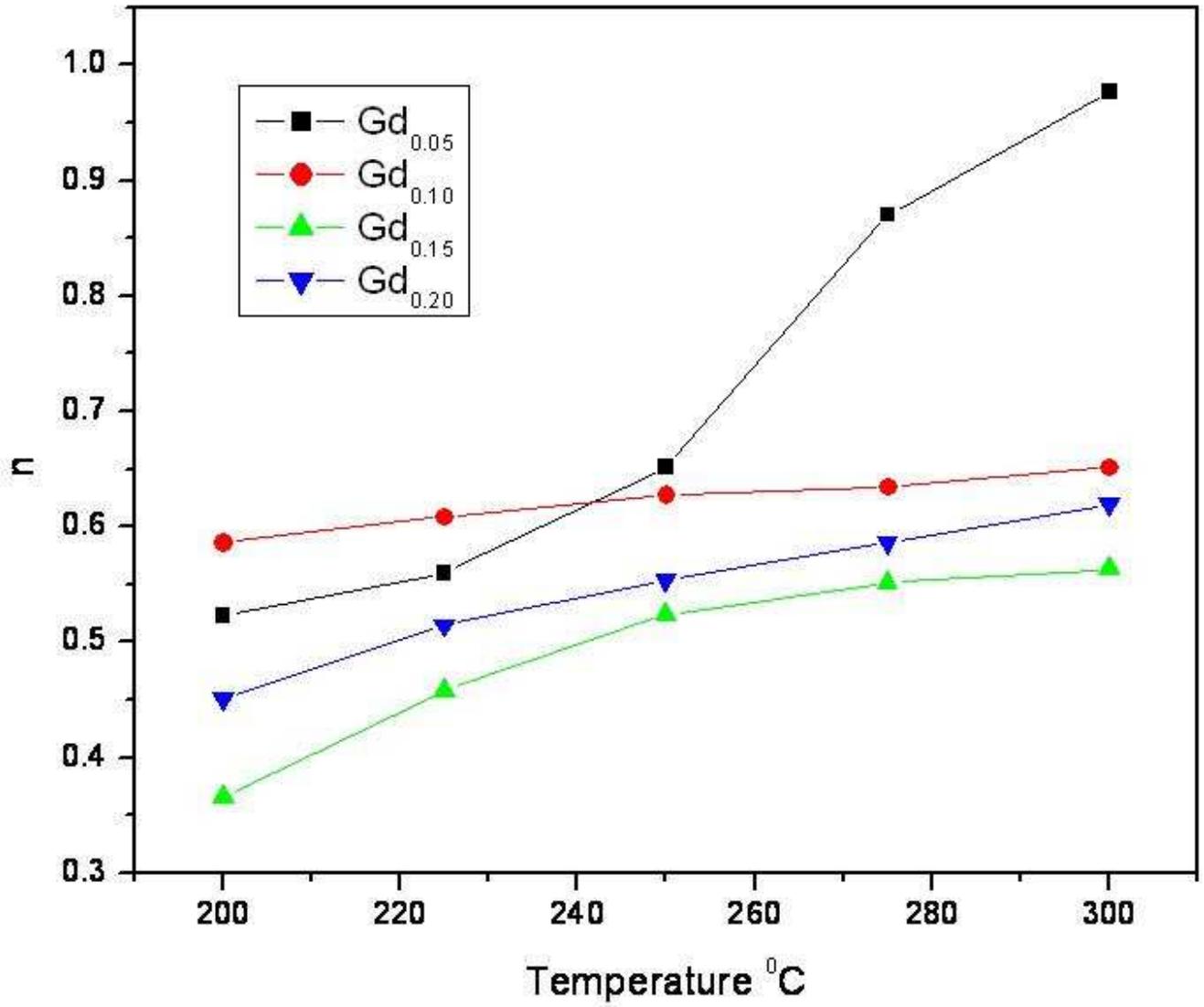

Figure. 9 Variation of A and n with temperature

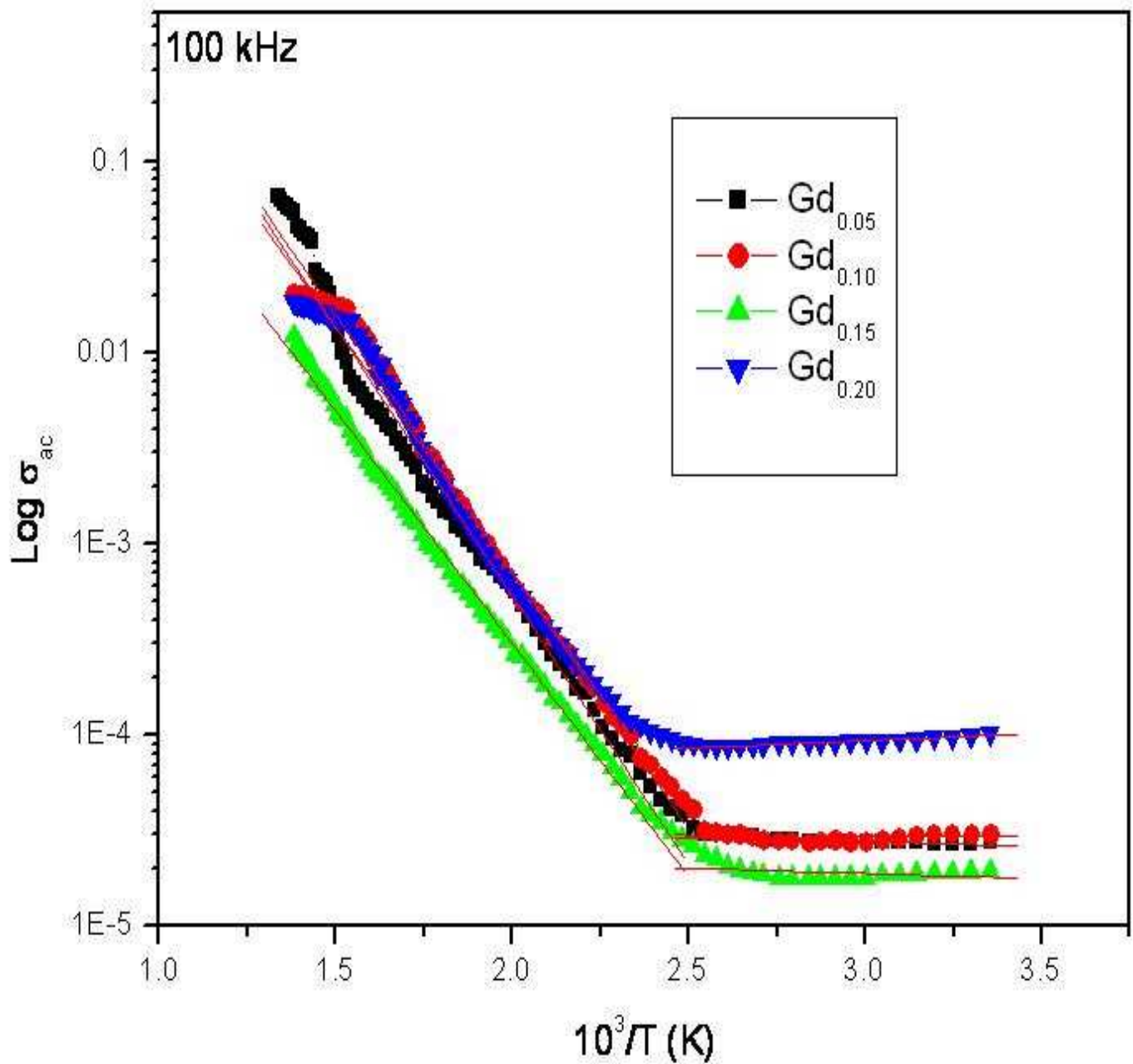

Figure. 10. Variation of ac conductivity ($\sigma_{ac}$) with ($10^3/T$) of $0.5(BiGd_xFe_{1-x}O_3)$-$0.5(PbZrO_3)$ (x=0.05, 0.10, 0.15, 0.20) at 100 kHz.

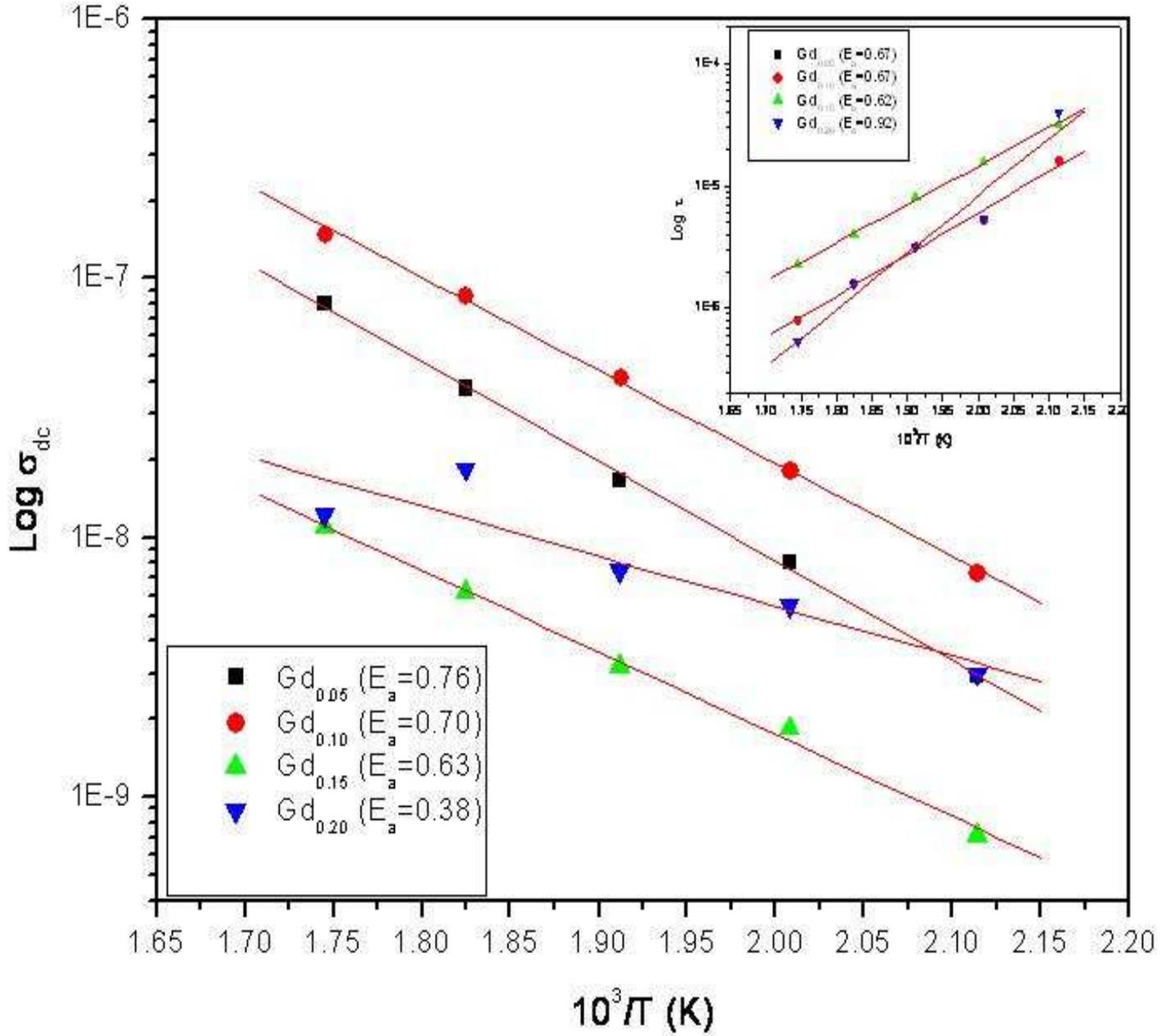

Figure. 11. Variation of $\sigma_{dc}$ and $\tau$ (inset) with $(10^3/T)$ of $0.5(BiGd_xFe_{1-x}O_3)$-$0.5(PbZrO_3)$ (x=0.05, 0.10, 0.15, 0.20).

Table 1 Summarizing of fitting parameters corresponding to equivalent circuits at $200^0C$ of Fig. 7.

| Different Parameters | X=0.05 | X=0.10 | X=0.15 | X=0.20 |
|---|---|---|---|---|
| $R_1$ | 1E-7 | 9.741E-5 | 63.4 | 177.3 |
| $R_2$ | 2.013E-4 | 9624 | 1.042E-5 | 4.092E-4 |
| $R_3$ | 1.131E-5 | 8.825E-4 | 3.901E-5 | 1.335E-5 |
| $R_4$ | | | 1.365E-5 | 8.1763E-4 |
| $C_1$ | 1.061E-10 | 1.981E-10 | 3.473E-11 | 4.016E-11 |
| $C_2$ | | | 2.297E-9 | 6.354E-10 |
| CPE | 1.053E-9 | 7.518E-10 | 2.609E-9 | 2.814E-8 |
| Frequency power (n) | 0.8303 | 0.8625 | 0.6362 | 0.4904 |
| Chi square | 0.02011 | 0.02124 | 0.006085 | 0.007499 |

Table 2 Fitting parameters obtained from the Jonscher power law at different temperatures. Comparisons between dc conductivity obtained from the fitted parameters and from Nyquist plots (parenthesis).

| T(°C) | x=0.05 | | x=0.10 | | x=0.15 | | x=0.20 | |
|---|---|---|---|---|---|---|---|---|
| | $\sigma_{dc}$ ($\Omega^{-1}m^{-1}$) | n | $\sigma_{dc}$ ($\Omega^{-1}m^{-1}$) | n | $\sigma_{dc}$ ($\Omega^{-1}m^{-1}$) | n | $\sigma_{dc}$ ($\Omega^{-1}m^{-1}$) | n |
| 200 | 0.00015 | 0.52254 | 0.00015 | 0.58654 | 0.00003 | 0.36544 | 0.00008 | 0.45089 |
| 225 | 0.00044 | 0.55945 | 0.00036 | 0.60866 | 0.0001 | 0.45851 | 0.00031 | 0.51484 |
| 250 | 0.001 | 0.65143 | 0.00073 | 0.62765 | 0.00022 | 0.5241 | 0.00044 | 0.55329 |
| 275 | 0.0025 | 0.87011 | 0.00154 | 0.63479 | 0.00042 | 0.55205 | 0.0011 | 0.58612 |
| 300 | 0.0048 | 0.97651 | 0.00296 | 0.65129 | 0.00073 | 0.5633 | 0.00199 | 0.6192 |

Table 3 Fitting parameters obtained from the comparisons between ac conductivity and inverse of temperature.

| Compositions (Gd$_x$) | Low temperature 25-120$^0$C Activation Energy (eV) | High temperature 145-445$^0$C Activation Energy (eV) |
|---|---|---|
| X=0.05 | 0.01216192 | 0.556769 |
| X=0.10 | 0.0151776 | 0.554254 |
| X=0.15 | 0.0105925 | 0.4820028 |
| X=0.20 | 0.0154037 | 0.5074238 |

Table 4 least-square straight-line fitting of the data to calculate the activation energy from the fig. 10

| Compositions ($Gd_x$) | DC Conductivity Activation Energy (eV) | Relaxation Time Activation Energy (eV) |
|---|---|---|
| X=0.05 | 0.76 | 0.67 |
| X=0.10 | 0.70 | 0.67 |
| X=0.15 | 0.63 | 0.62 |
| X=0.20 | 0.38 | 0.92 |